\documentclass[12pt]{article}
\usepackage{subeqnar,epsf}

\begin{document}

\title{Four-Loop Perturbative Expansion for the Lattice $N$-Vector Model}
\author{
  \\
  {\small Sergio Caracciolo}              \\[-0.2cm]
  {\small\it Dipartimento di Fisica and INFN -- Sezione di Lecce}  \\[-0.2cm]
  {\small\it Universit\`a degli Studi di Lecce}        \\[-0.2cm]
  {\small\it I-73100 Lecce, ITALIA}          \\[-0.2cm]
  {\small Internet: {\tt CARACCIO@UX1SNS.SNS.IT}}     \\[-0.2cm]
  \\[-0.1cm]  \and
  {\small Andrea Pelissetto}              \\[-0.2cm]
  {\small\it Dipartimento di Fisica and INFN -- Sezione di Pisa}    \\[-0.2cm]
  {\small\it Universit\`a degli Studi di Pisa}        \\[-0.2cm]
  {\small\it I-56100 Pisa , ITALIA}          \\[-0.2cm]
  {\small Internet: {\tt PELISSET@SUNTHPI1.DIFI.UNIPI.IT}}   \\[-0.2cm]
  {\protect\makebox[5in]{\quad}}  
  \\
}
\vspace{0.5cm}

\maketitle
\thispagestyle{empty}   

\vspace{0.2cm}

\begin{abstract}
We compute the four-loop contributions to the $\beta$-function and
the anomalous dimension of the field for the $O(N)$-invariant
$N$-vector model.  These results are used
to compute the second analytic corrections to
the correlation length and the general spin-$n$ susceptibility.
\end{abstract}

\clearpage

\newcommand{\be}{\begin{equation}}
\newcommand{\ee}{\end{equation}}
\newcommand{\<}{\langle}
\renewcommand{\>}{\rangle}

\def\spose#1{\hbox to 0pt{#1\hss}}
\def\ltapprox{\mathrel{\spose{\lower 3pt\hbox{$\mathchar"218$}}
 \raise 2.0pt\hbox{$\mathchar"13C$}}}
\def\gtapprox{\mathrel{\spose{\lower 3pt\hbox{$\mathchar"218$}}
 \raise 2.0pt\hbox{$\mathchar"13E$}}}

\def\bsigma{\mbox{\protect\boldmath $\sigma$}}
\def\bpi{\mbox{\protect\boldmath $\pi$}}
\def\btau{\mbox{\protect\boldmath $\tau$}}
\def\hatp{\hat p}
\def\hatl{\hat l}

\def\msbar{ {\overline{\hbox{\scriptsize MS}}} }
\def\normalmsbar{ {\overline{\hbox{\normalsize MS}}} }

\newcommand{\R}{\hbox{{\rm I}\kern-.2em\hbox{\rm R}}}

\newcommand{\reff}[1]{(\ref{#1})}

\section{Introduction}

Non-linear $\sigma$-models have been and are being investigated in
theoretical physics for a variety of reasons:
in condensed matter physics the non-linear $N$-vector model
describes the critical behaviour of systems with an
$N$-component order parameter~\cite{Kadanoff,Fisher}; in elementary
particle physics two-dimensional $\sigma$-models serve  as
a playground for testing ideas which are relevant to four
dimensional gauge theories: indeed they are asymptotically
free and can be studied with a weak-coupling perturbative expansion
\cite{Polyakov_75,Brezin_76,Bardeen_76,Dadda}.

The simplest example is the $\sigma$-model where the fields
take values in the sphere $S^{N-1}$
and where the action is invariant under global $O(N)$ transformations.
Besides perturbation theory, it can be studied using different techniques.
It can be solved in the $N=\infty$ limit~\cite{Stanley,DiVecchia}
and $1/N$ corrections can be systematically computed~\cite{Muller,Flyvbjerg,
Campostrini_90ab}.
Moreover an exact $S$-matrix can be computed~\cite{Zamolodchikov_79,
Polyakov-Wiegmann_83}
and, using the thermodynamic Bethe ansatz , the exact mass-gap
of the theory in the limit $\beta\to\infty$ can been obtained
\cite{Hasenfratz-Niedermayer_1,Hasenfratz-Niedermayer_2,Hasenfratz-Niedermayer_3}.
The model has also been studied numerically:
extensive simulations have been performed
for $N=3$ \cite{CEFPS,CEPS_O3_FSS},
$N=4$ \cite{Wolff_O4_O8,MGMC_O4}
and $N=8$ \cite{Wolff_O4_O8}.
The results for the correlation length agree with the conventional predictions
--- including the nonperturbative coefficient ---
to within about 4\% for $N=3$ (at $\xi \approx 10^5$),
6\% for $N=4$
(for $\xi \approx 100$) and 1\% for $N=8$ (for $\xi \approx 30$).
The remaining deviations are not much larger than the three-loop
correction: for $N=3$ (resp. 4, 8), at the largest $\beta$ where
Monte Carlo data are available, the three-loop correction is about
3\% (resp. 2\%, 0.5\%). For this reason we expect the inclusion of
the four-loop term to improve sensibly the agreement with the conventional
predictions.

In this paper we compute the $\beta$-function and the anomalous
dimension of the field up to four-loops, thus extending previous
work by Falcioni and Treves \cite{Falcioni-Treves}. From this computation
we obtain the second analytic coefficients in the perturbative
expansion of the correlation length $\xi$ and of the vector susceptibility
$\chi$
and the third analytic correction to the ratio $\chi/\xi^2$.
Using results obtained in \cite{CP-3loop} we can also compute
the second correction to the general spin-$n$ susceptibility.
Some technical details concerning the computation are reported in Appendix A.
A check of the results is provided by the $1/N$ results of
\cite{Campostrini_90ab}: in Appendix B we have checked the correctness of the
large-$N$ limit of our results.
We have finally compared our four-loop prediction to the
available data for the correlation length (a much more detailed
comparison for $N=3$, which includes also the susceptibilities,
will appear in \cite{CEPS_93}): we find that the discrepancy between
theory and experiment at the largest $\beta$ today available is now
reduced to 2\%, 4\%, 0.5\% for $N=3,4,8$, the four-loop correction
being of order 2\%, 2\%, 0.5\% in the three cases. The remaining difference
should be ascribed to higher-loop corrections: for $N=8$,
using the large-$N$ results, we have indeed verified that, if all analytic
corrections were included up to eight loops, the discrepancy
should be $\ltapprox 0.1$\%.

\section{Four-loop RG Functions} \label{sec2}

In this paper we consider the nearest-neighbor lattice $N$-vector model
in two dimensions. The fields
are unit-length spins $\bsigma_x\in R^N$ and the hamiltonian is given by
\be
{\cal H}^{latt} \, =\, - \sum_{x\mu} \bsigma_x \cdot \bsigma_{x + \mu}
\ee
The partition function is given by
\be
Z\, =\, \int \exp (-\beta {\cal H}^{latt})\, \prod_x d^N\bsigma_x  \,
                \delta (\bsigma_x^2 - 1)
\ee
As it is well known, the perturbative expansion of this model in
two dimensions is plagued by infrared
divergences. We will not discuss this problem here and we will adopt
the common technique of adding a magnetic field $h$ to the
hamiltonian as an infrared regulator. Thus, if the magnetic field points in
the first direction we have the hamiltonian
\be
{\cal H}^{latt} \, =\, - \sum_{x\mu} \bsigma_x \cdot \bsigma_{x + \mu}-\,
    h\, \sum_x \bsigma_x^1
\ee
The perturbative expansion is then obtained by considering small
fluctuations around the direction of the magnetic field. Thus one sets
\be
    \bsigma_x \, =\, \left( \sqrt{1 - \bpi^2_x},\bpi_x\right)
\ee
and expands the Hamiltonian in powers of $\bpi$.

We will now compute the four-loop $\beta$-function and anomalous
dimension of the field $\bsigma$.
In principle this can be done
through a direct lattice computation. However it is much simpler to
take advantage of the fact that the four-loop calculation has already been done
for the continuum theory in dimensional regularization
\cite{Brezin-Hikami,Hikami,Wegner_1}. This allows us to
compute the four-loop contribution by performing a lattice computation at
three loops. The idea is to compute the finite renormalization constants
$Z_1(\beta,\mu a)$ and $Z_2(\beta, \mu a)$ which relate the
Green's functions in the $\overline{MS}$-scheme and on the lattice.

More precisely, define
$\Gamma^{(n)}_{latt}(p_1,\ldots, p_n;\beta,h;1/a)$ as the lattice
$n$-point one-particle-irreducible correlation function for the $\bpi$-field
and $\Gamma^{(n)}_{\overline{MS}}(p_1,\ldots, p_n;\beta,h;\mu)$ its
counterpart in the $\overline{MS}$ scheme. Then the general results
of~\cite{Brezin-LeGuillou-ZinnJustin} imply
\be
\Gamma^{(n)}_{latt} (p_1,\ldots,p_n;\beta,h;1/a)\, =\, Z_2^{n/2}\,
\Gamma^{(n)}_{\overline{MS}} (p_1,\ldots,p_n;Z_1^{-1}\beta,Z_1 Z_2^{-1/2} h;\mu
)
\label{latticeMS}
\ee
It follows that the $\beta$-functions $W(\beta)$ and the anomalous dimensions
$\gamma(\beta)$ in the two schemes are related by
\begin{eqnarray}
W^\msbar(Z_1^{-1} \beta) &=& W^{latt} (\beta)\left( Z_1 + {1\over\beta}
{\partial Z_1\over\partial\beta^{-1}}\right) \label{eq2.4}\\
\gamma^\msbar(Z_1^{-1} \beta) &=& \gamma^{latt}(\beta) - \,
W^{latt} (\beta) {1\over Z_2} \,{\partial Z_2\over\partial\beta^{-1}}
\label{eq2.5}
\end{eqnarray}
In general we expand, on the lattice as well as in the $\msbar$ scheme,
\be
W^{scheme}(\beta)\, =\,
- {w_0\over\beta^2} - {w_1\over\beta^3} -
{w_2^{scheme} \over\beta^4} -
{w_3^{scheme} \over\beta^5}
+\, O(\beta^{-6})
\ee
and
\be
\gamma^{scheme}(\beta)\, =\,
 {\gamma_0\over\beta} + {\gamma_1^{scheme}\over\beta^2} +
{\gamma_2^{scheme} \over\beta^3} +
{\gamma_3^{scheme} \over\beta^4}
+\, O(\beta^{-5})
\ee
The coefficients
$w_0$, $w_1$ and $\gamma_0$ are universal
in the sense that they do not depend on the renormalization procedure
and for this reason we have not added the superscript $scheme$.
They are explicitly given by
\begin{eqnarray}
w_0\, &=&\, {N-2\over 2\pi} \\
w_1\, &=&\, {N-2\over (2\pi)^2} \\
\gamma_0 &=& {N-1\over 2\pi}
\end{eqnarray}
All other terms instead are scheme-dependent. In $\msbar$ they are
explicitly given by~\cite{Brezin-Hikami,
Hikami,Wegner_1,Wegner_2}
\begin{eqnarray}
w_2^\msbar & =& {1\over4} {N^2-4\over (2\pi)^3} \\
w_3^\msbar & =& {N-2\over (2\pi)^4}\left[-{1\over12}(N^2-22N+34)
+{3\over2} \zeta(3) (N-3)\right]  \\
\gamma_1^\msbar &=& 0 \\
\gamma_2^\msbar &=& {3\over32\pi^3}(N-1)(N-2) \\
\gamma_3^\msbar &=& {1\over 192\pi^4}(N-1)(N-2)[4(5-N)+3(3-N)\zeta(3)]
\end{eqnarray}
where $\zeta(3)\approx1.2020569$.
On the lattice we have~\cite{Falcioni-Treves,CP-3loop}
\be
w^{latt}_2 =
   {N-2\over (2\pi)^3}\left[\left({1\over2} + {\pi^2\over8} - 4\pi^2 G_1
    \right)
   (N-2) + 1 + {\pi\over2} - {5\pi^2\over24}\right]
\ee
and
\begin{eqnarray}
\gamma_1^{latt} &=& \, {N-1\over8\pi} \\
\gamma_2^{latt} &=& \, {N-1\over (2\pi)^3}\left[
\left({1\over2} - {\pi^2\over8} + 4 \pi^2 G_1 \right) (N-2)
        \, +\, {11\pi^2\over24}
\right]
\end{eqnarray}
where $G_1 \approx 0.0461636$.

\begin{figure}
\vspace*{0cm} \hspace*{-0cm}
\begin{center}
\epsfxsize = \textwidth
\leavevmode\epsffile{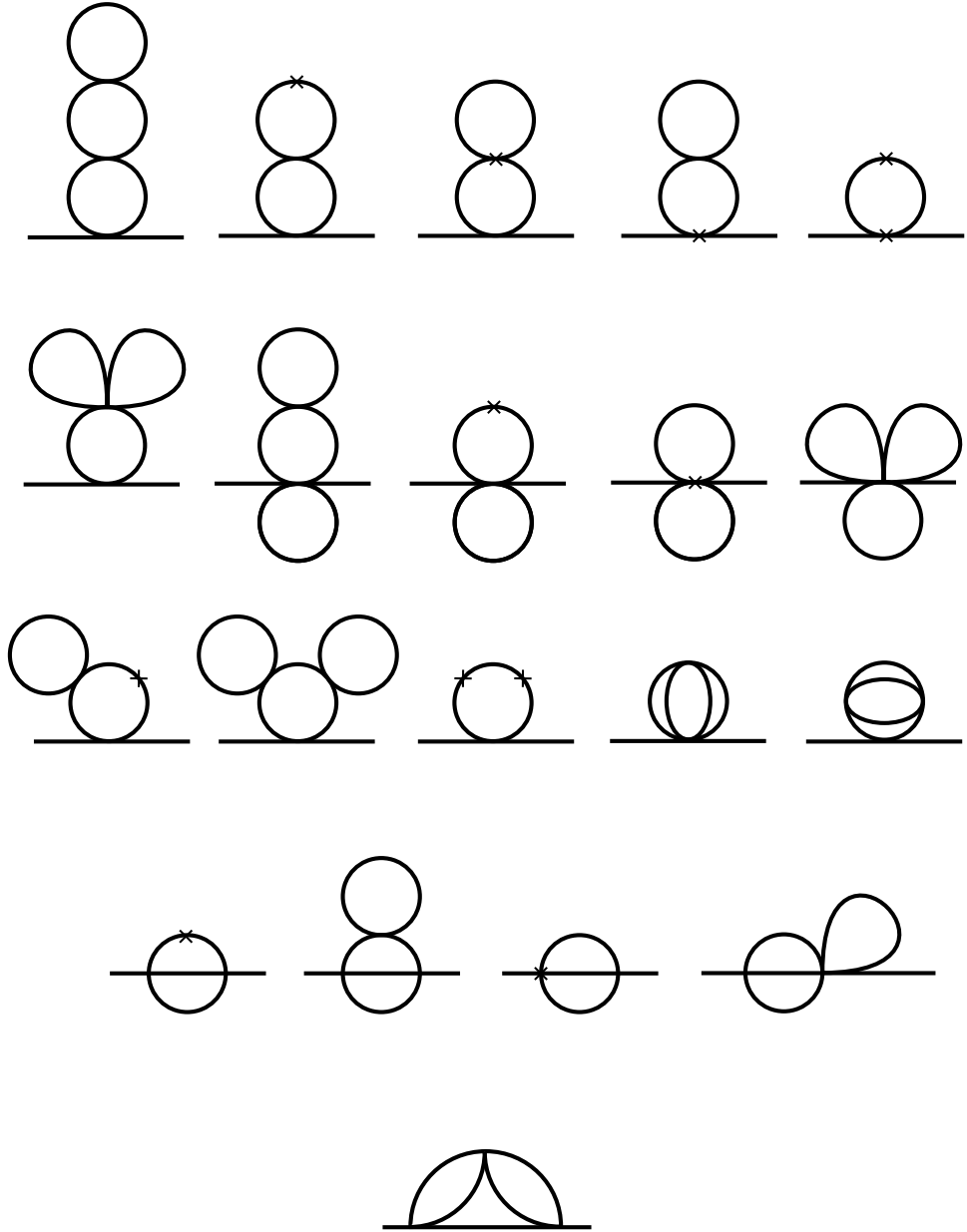}
\end{center}
\caption{Feynman graphs appearing in the computation of the two-point
function at three-loops.
}
\label{fig1}
\end{figure}

We will now compute $w^{latt}_3$ and $\gamma_3^{latt}$.
We must first of all compute the three-loop self-energy on the lattice
for the $\bpi$-field. The Feynman graphs are reported in Fig. \reff{fig1}.
We get
\begin{eqnarray}
&& (N-1) h
      \, \left\{
       {1\over 3072 \pi^3} (3N-5) (5N-9) \log^3 {h a^2\over 32} \right.
      \nonumber \\
 &&\qquad + {1\over 512 \pi^3} \left[ 4N^2 - 23 N + 29 - 2 \pi (3 N -5)\right]
                  \log^2 {h a^2\over 32} \nonumber \\
 &&\qquad - {1\over 4\pi} \left[ {1\over192} (3 N - 17) +
                        {1\over 64\pi} (5 N - 11) -
              {1\over128\pi^2} (N-3)^2 - \right. \nonumber \\
 && \qquad\qquad \left. \left. {1\over 2} G_1 (N-2)
          \right] \log{h a^2\over 32} +
       A_3 \right\} \, + \nonumber \\
 &&   p^2 \, \left\{
       {1\over384\pi^3} (2 N -3) (N-1) \log^3 {h a^2\over 32} + \right.
        \nonumber \\
 &&\qquad + {1\over256\pi^3} \left[ 11 N^2 - 41 N + 36 - 2 \pi (N-1)\right]
                 \log^2 {h a^2\over 32}  \nonumber \\
 &&\qquad - {1\over 4\pi}
       \left [ - {1\over64\pi^2} (5 N^2 - 42 N + 65) + {1\over32\pi}(7 N -13)
               - \right. \nonumber \\
 && \qquad\qquad \left.\left. R (2 N -3) (N-2)\vphantom{{1\over2}}\right]
           \log{h a^2\over 32} + B_3 \right\}
\end{eqnarray}
where
\begin{eqnarray}
A_3 &=& {1\over 768\pi} (3 N + 8) - {1\over 256\pi^2} (N-3) +
       {7\over3072\pi^3} (11 N -21) \zeta(3) \nonumber \\
  && \qquad - {L_1\over12} +\, {N-1\over24} V_1 +\, {N-2\over8} V_2 \\[3mm]
B_3 &=& - {2 N -7\over 192} +\, {3 N + 2\over 384\pi} -\,
         {5 N -11\over 128 \pi^2} + {1\over1536\pi^3} (98 N^2 - 557 N + 729)
         \zeta(3) \nonumber \\
  && \qquad - {N-6\over48} J - {L_1\over6} + {1\over12} (N^2 - 8N + 13) V_1 +
     {1\over 16} (N-2)(N-3) (2 V_3 - K) \nonumber \\
  && \qquad + {N-2\over4} (3 G_1 - 3 R + V_2
     - 4 V_4 + 2 V_5 + 2 V_6 + 16 W_2)
\end{eqnarray}
Here $V_1$, $\ldots, V_6$, $G_1$, $R$, $J$ and $W_2$ are finite lattice
integrals. Their definition is reported in Appendix A.1 and their
numerical value in table~\ref{tavola1}.

Most of the graphs can be computed with a limited effort:
using the table of integrals appearing in Appendix A.2, all graphs but
the last one in Fig. \reff{fig1}
have been reduced automatically using the symbolic language
{\sc mathematica}. The computation of the last graph was much more difficult
and involved. The whole computation has been done independently by the
two authors, and many intermediate results have been checked numerically.
Some technical details can be found in Appendix A.
\begin{table}
\begin{center}
\tabcolsep 6pt      
\doublerulesep 2pt  
\begin{tabular}{|c|c|}
\hline
\multicolumn{2}{|c|}{Constants} \\
\hline
$G_1$ & \hphantom{$-$} 0.0461636 \\
$R $ & \hphantom{$-$} 0.0148430 \\
$J $ & \hphantom{$-$} 0.1366198 \\
$K $ & \hphantom{$-$} 0.095887\hphantom{0}  \\
$L_1$ & \hphantom{$-$} 0.0029334 \\
$V_1$ & \hphantom{$-$} 0.016961\hphantom{0} \\
$V_2$ & $-$ 0.00114\hphantom{00} \\
$V_3$ & \hphantom{$-$} 0.07243\hphantom{00}\\
$V_4$ & $-$ 0.0013125 \\
$V_5$ & \hphantom{$-$} 0.010063\hphantom{0} \\
$V_6$ & \hphantom{$-$} 0.017507\hphantom{0} \\
$W_1$ & $-$ 0.0296860 \\
$W_2$ & \hphantom{$-$} 0.00221\hphantom{00} \\
\hline
\end{tabular}
\end{center}
\caption{
Numerical value of the lattice integrals appearing in the $\beta$ and
$\gamma$ functions. Their explicit definitions are reported in Appendix A.1.
}
\label{tavola1}
\end{table}

We must also compute the same correlation function in the continuum theory.
In the $\msbar$-scheme we get
\begin{eqnarray}
 h (N-1) && \!\!\!\!\!\!\! \left[ {1\over 3072\pi^3} (5 N -9) (3N-5)
          \log^3 {h\over \mu^2} +
          {1\over 512\pi^3} (4 N^2 - 23 N + 29) \log^2 {h\over \mu^2}
          \right. \nonumber \\
         && \left.
         + {1\over 512\pi^3} (N-1)^2 \log {h\over \mu^2} +\,
          {7\over 3072\pi^3} (11 N - 21) \zeta(3) \right] \\
+ p^2 && \!\!\!\!\!\!\! \left[
        {1\over 384\pi^3} (2 N -3) (N-1) \log^3 {h\over \mu^2} +
        {1\over 256\pi^3} (11 N^2 - 41 N + 36) \log^2 {h\over \mu^2} +
          \right. \nonumber \\
         &&
        + {1\over 256\pi^3} (5 N^2 - 42 N + 65) \log {h\over \mu^2} +
          {1\over 4 \pi} R (2 N -3) (N-2) \log {h\over \mu^2} - \nonumber \\
        && \left. - {1\over 96\pi^3} (N-8) (N-2) -
             {3\over 2\pi} (N-2) R +
             {1\over 1536 \pi^3} (98 N^2 - 117 N - 151)\zeta(3)
             \right] \nonumber
\end{eqnarray}
{}From these expressions it is easy to obtain the renormalization constants
$Z_1$ and $Z_2$. We expand
\begin{eqnarray}
Z_1 &=& 1 + {Z_{11}\over \beta} + {Z_{12}\over \beta^2} +
            {Z_{13}\over \beta^3} +\, O(\beta^{-4}) \\
Z_2 &=& 1 + {Z_{21}\over \beta} + {Z_{22}\over \beta^2} +
            {Z_{23}\over \beta^3} +\, O(\beta^{-4})
\end{eqnarray}
The terms proportional to $\beta^{-1}$ and $\beta^{-2}$ have
already been computed in \cite{CP-3loop}.
For the three-loop terms we get
\begin{eqnarray}
Z_{13} &=& - {(N-2)^3\over 64 \pi^3} \log^3 {a^2\mu^2\over32}
+\, {(N-2)^2\over 64 \pi^3} (5 + 3 \pi) \log^2 {a^2\mu^2\over32}
\nonumber \\
&& \quad +\, {N-2\over 192 \pi^3}
  \left [ 3 (N-6) - 18 \pi + (3 N -20) \pi^2 - 96 \pi^2 G_1 (N-2) \right]
  \log {a^2\mu^2\over32} \nonumber \\
&& \quad - {5 N -21\over 192} - {(N-2)^2\over 64 \pi} -
   {3 (N-2)\over 64 \pi^2} - {1\over 192\pi^3} (N-2)(N+10)  \nonumber \\
&& \quad - {55\over 192 \pi^3} (N-2) \zeta(3)
 - {N-6\over 48} J + {1\over 4\pi} (N-2)(N-2 + 5\pi) G_1 \nonumber \\
&& \quad + {N-2\over 2\pi} (3-\pi) R - {1\over 16}(N-2)(N-3)(K-2 V_3)
   \nonumber \\
&& \quad + {N-2\over 4} \left[{2\over3} L_1 - 2 V_1 -
   (N-2) V_2 - 4 V_4 + 2 V_5 + 2 V_6 + 16 W_2 \right] \\
Z_{23} &=& (N-1) \left\{ - {1\over 384 \pi^3} (3 N - 5) (2 N -3)
               \log^3 {a^2\mu^2\over32} \right. \nonumber \\
&& \quad +\,
  {1\over 64 \pi^3} [N-2 + \pi(2 N - 3)] \log^2 {a^2\mu^2\over32}
    \nonumber \\
&& \quad +\, {1\over 384\pi^3} \left[
    -12 (N-2) + (3 N - 17) \pi^2 - 96 \pi^2 G_1 (N-2)\right]
     \log {a^2\mu^2\over32}
   \nonumber \\
&& \quad \left. - {N-2\over 64 \pi^3} (1 + \pi^2) +
     {L_1\over 6} - {N-1\over 12} V_1 + {N-2\over 4 \pi} (G_1 - \pi V_2)
     \right \}
\end{eqnarray}
Then, using \reff{eq2.4} and \reff{eq2.5}, we finally obtain
\begin{eqnarray}
w_3^{latt} &=& {N-2\over 2\pi} \left\{ {2 N -7\over 96} +
      {N^2 - 4 N + 5\over 32 \pi} + {N\over 16\pi^2} +
      {5 N - 9\over 16 \pi^3} + {73 N - 164\over 96 \pi^3} \zeta(3)
      \right. \nonumber \\
     && \quad - {N-2\over 2\pi} (N-2 + 3\pi) G_1 - {N-2\over \pi} (3-\pi) R
     + {N-6\over 24} J - {N-2\over 3} L_1 \nonumber \\
     && \quad + {1\over 8} (N-3)(N-2)(K-2 V_3) \nonumber \\
     && \quad \left. +
       {N-2\over2} (2 V_1 + (N-2) V_2 + 4 V_4 - 2 V_5 - 2 V_6 - 16 W_2)
       \right\}
 \\
\gamma_3^{latt} &=& {N-1\over 2\pi} \left\{ - {5 N -21\over 192} +
      {(N-2)^2\over 32 \pi} + {3(N-2)\over 32\pi^2} +
      {N - 2\over 16 \pi^3} \right. \nonumber \\
     && \quad - {N - 2\over 192 \pi^3}\, (6 N + 37) \zeta(3)
     - {N-2\over 4\pi} \left[2(N-2) - 5\pi\right] G_1
              + {N-2\over 2\pi} (3-\pi) R  \nonumber \\
     && \quad - {N-6\over 48} J - {N-2\over 3} L_1
        - {1\over 16} (N-3)(N-2)(K-2 V_3) \nonumber \\
     && \quad \left. + {N-2\over4}
    \left[(N-3) V_1 + 2 (N-2) V_2 - 4 V_4 + 2 V_5 + 2 V_6 + 16 W_2\right]
       \right\}
\end{eqnarray}
A check of these results is provided by the $1/N$-results of
\cite{Campostrini_90ab}. In the large-$N$ limit we get
from the previous expressions
\begin{eqnarray}
{w_3^{latt}\over N^3} &=& {1\over 2\pi^2} \left[ {1\over32} - {G_1\over2}
     + {\pi\over8} \left( K + 4 V_2 - 2 V_3\right)\right] +\, O(1/N)
  \label{eq2.31} \\
{\gamma_3^{latt}\over N^3} &=& {1\over2\pi}\left[ {1\over32\pi} -
     {1\over 32\pi^3}\zeta(3) - {G_1\over 2\pi} - {K\over16} +
     {V_1\over4} + {V_2\over2} + {V_3\over8}\right] +\, O(1/N)
  \label{eq2.32}
\end{eqnarray}
In Appendix B we have checked that \reff{eq2.31}/\reff{eq2.32}
agree with the predictions of the $1/N$ expansion.

\section{Long-Distance Quantities} \label{sec3}

We will now use $w_3^{latt}$ and $\gamma_3^{latt}$ to compute the second
analytic correction to the correlation length $\xi$ and
spin susceptibility $\chi$.

Let us begin with $\xi$. In general we have
\begin{eqnarray}
\xi &=& C_{\xi} e^{\beta/w_0}\left( {w_0\over \beta}\right)^{w_1/w_0^2}
       \exp\left[ \int^{1/\beta}_0 dt
       \left( {1\over W^{latt}(1/t)} + {1\over w_0 t^2} -
              {w_1\over w_0^2 t}\right)\right] \nonumber \\
&=& C_{\xi}
\left({N-2\over 2\pi\beta}\right)^{1/(N-2)}
\exp\left( {2\pi\beta\over N-2}\right) \left(1 +
\sum_{n=1}^\infty{a_n\over\beta^n} \right)
\label{eq2.90}
\end{eqnarray}
The constant $C_{\xi}$ is non-perturbative and its value depends
on the explicit definition of the correlation length.
For the isovector exponential correlation length $\xi_V^{(exp)}$, which
controls the large-distance behaviour of the two-point
function $\<\bsigma_0\cdot\bsigma_x\>$,
an explicit expression has been obtained
using the thermodynamic Bethe ansatz
\cite{Hasenfratz-Niedermayer_1,Hasenfratz-Niedermayer_2,Hasenfratz-Niedermayer_3}.
Explicitly
\be
C_{\xi_V^{(exp)}} = \left( {e\over 8} \right)^{1/(N-2)}
\Gamma\left(1+{1\over N-2}\right) 2^{-5/2} \exp\left[-{\pi\over 2(N-2)}\right]
\label{Hasenfratzconst}
\ee
Other possibilities are the second-moment correlation length
\be
     \xi^{(2)}_V \, =\, \sqrt{\sum_x |x|^2 \< \bsigma_0\cdot\bsigma_x\> \over
                              \sum_x \< \bsigma_0\cdot\bsigma_x\>}
\ee
in the isovector channel, or the analogous quantities in higher-isospin
channels. For instance, in studies of mixed $O(N)$/$RP^{N-1}$
\cite{CEPS_93},
the isospin-2 correlation lengths associated with
the correlation $\<(\bsigma_0\cdot\bsigma_x)^2 - 1/N\>$
were introduced: the exponential
isotensor correlation length $\xi_T^{(exp)}$ and the corresponding
second-moment correlation length $\xi_T^{(2)}$. Using the
fact that in the $O(N)$ $\sigma$-model no bound states exist
\cite{Zamolodchikov_79}, we have immediately
\be
      C_{\xi_T^{(exp)}} = {1\over2}\, C_{\xi_V^{(exp)}}
\ee
For the second-moment correlation lengths no exact expressions exist.
However, in the large-$N$ limit we get \cite{CP-unosuN}
\begin{eqnarray}
 C_{\xi_V^{(2)}} &=&   C_{\xi_V^{(exp)}} \left(1 - {0.0032\over N} + O(1/N^2)
    \right) \\
 C_{\xi_T^{(2)}} &=&{1\over \sqrt{6}} \,
                       C_{\xi_V^{(exp)}} \left(1 - {1.2031\over N} + O(1/N^2)
    \right)
\end{eqnarray}

Let us now consider the perturbative corrections to the universal behaviour.
The first coefficient $a_1$ was computed in \cite{Falcioni-Treves}:
\be
a_1 =\, {1\over N-2} \left[\left( {1\over 4\pi} + {\pi \over 16} -
        {2 \pi G_1}\right) (N-2) + {1\over 4} - {5 \pi\over 48}\right]
\ee
We can now compute $a_2$ to get
\begin{eqnarray}
a_2 &= & {1\over 192(N-2)^2} \left[
    {\pi^2\over 24} (3 N - 11)^2 + \pi (4 N^2 - 19 N + 17) +
    6 N^3 - 39 N^2 + 95 N - 76\right] \nonumber \\
  &&\quad + {1\over 32 \pi^2 (N-2) } \left[ 7N-16 + 2 \pi (N-1)\right] +\,
    {1\over 96 \pi^2} {73 N - 164\over N-2}\, \zeta(3) \nonumber \\
 && \quad + {G_1\over 24 (N-2)} \left[ -12 (N-2) (N-3) - 12 \pi (3 N - 5)
      - (3N-11)\pi^2 \right] + 2 \pi^2 G_1^2 \nonumber \\
 && \quad - (3-\pi)\, R +\, {\pi\over 24}\, {N-6\over N-2}\, J
     - {\pi\over3} L_1 +\,
     {\pi\over8} (N-3) (K - 2 V_3) \nonumber \\[3mm]
 && \quad + {\pi\over2} \,
     \left( 2 V_1 + (N-2) V_2 + 4 V_4 - 2 V_5 - 2 V_6 -16 W_2
    \right)
\end{eqnarray}
Analogous expressions can be derived for the isovector susceptibility
$\chi_V = \sum_x \< \bsigma_0\cdot \bsigma_x\>$. We get
\begin{eqnarray}
\chi_V &=& C_\chi e^{2\beta/w_0}
\left( {w_0\over \beta}\right)^{2 w_1/w_0^2+\gamma_0/w_0} \,\times \nonumber \\
&& \quad \exp\left[\int_0^{1/\beta} dx\, \left({2\over W^{latt}(1/x)} +
\, {2\over w_0 x^2} -\, {2w_1\over w_0^2 x}-\,
{\gamma^{latt}(1/x)\over W^{latt}(1/x)}
- \, {\gamma_0\over w_0 x}\right)\right]
\nonumber \\
&=& C_\chi e^{4\pi\beta/(N-2)}
\left({2\pi\beta\over N-2}\right)^{-(N+1)/(N-2)}\,
\left\{1 + \sum_{n=1}^\infty {b_n\over\beta^n} \right\}
\label{eq2.131}
\end{eqnarray}
The (non-universal) constant $C_\chi$ cannot be computed in perturbation
theory,
and no exact expression is available at present.
We can evaluate $C_\chi$ in the large-$N$ limit.
Using the $1/N$ results of
\cite{Campostrini_90ab}
we obtain the following expression:
\begin{eqnarray}
C_\chi &=& {\pi\over 16}\, \left[1 + {1\over N} (4 + 3 \gamma_C
-\pi - 3 \gamma_E - 7 \log 2) +\, O(1/N^2)\right]
\nonumber \\
& \approx & 0.196 \, [ 1 - 4.267/N + \, O(1/N^2)]\,
\end{eqnarray}
where $\gamma_E\approx 0.5772157$ is Euler's constant and
\be
\gamma_C \, =\, \log\left({\Gamma(1/3)\Gamma(7/6)\over\Gamma(2/3)\Gamma(5/6)}
\right)\, \approx \, 0.4861007
\ee

We will also consider the ratio $\chi_V/\xi^2$ as in this case we can
compute an additional analytic correction. We write
\be
{\chi_V\over \xi^2 }\, =\, R_V
\left({2\pi\beta\over N-2}\right)^{-(N-1)/(N-2)} \left(1 + \,
\sum_{n=1}^\infty  {c_n\over \beta^n} \right)
\ee
where $R_V= C_\chi/C^2_\xi$ is a non-perturbative universal quantity.

The explicit values of $b_1$, $c_1$ and $c_2$ are reported in
\cite{Falcioni-Treves,CP-3loop}. Explicitly
\begin{eqnarray}
c_1 &=& {1\over 4\pi}\,{N-1\over N-2} ( \pi-2 ) \nonumber \\
c_2 &=& {N-1\over (N-2)^2}\, \left[ - {1\over 96} (3 N^2 - 23 N + 31)
       +{N-1\over 8\pi^2} - {2N-3\over 8\pi} + (N-2)^2 G_1\right] \nonumber\\
b_1
&= &{1\over 2\pi(N-2)} \left[-1 +{\pi\over2}(N+1) +
{\pi^2\over4}\left(N-{11\over3}\right) - 8\pi^2 (N-2)G_1 \right]
\end{eqnarray}
Here we will compute $b_2$ and $c_3$.
They are given by:
\begin{eqnarray}
b_2 &= & {1\over (N-2)^2} \left[
    {\pi^2\over 1152} (3 N - 11)^2 + {\pi\over96} (7 N^2 - 30 N + 17) +
    {1\over96} (N-1)(3 N^2 - 13 N + 51) \right. \nonumber \\
  &&\quad  \left. - {1\over 8\pi} (2 N -1) +
        {1\over 8\pi^2} (3 N^2 - 13 N + 15) \right] +\,
    {1\over 48 \pi^2} {73 N - 164\over N-2}\, \zeta(3) \nonumber \\
 && \quad + {G_1\over 6 (N-2)} \left[ 6(3N-4) - 6 \pi (4 N - 5)
      - (3N-11)\pi^2 \right] + 8 \pi^2 G_1^2 \nonumber \\
 && \quad - 2 (3-\pi)\, R +\, {\pi\over 12}\, {N-6\over N-2}\, J
     - {2\pi\over3} L_1 +\,
     {\pi\over4} (N-3) (K - 2 V_3) \nonumber \\[3mm]
 && \quad + \pi \,
     \left( 2 V_1 + (N-2) V_2 + 4 V_4 - 2 V_5 - 2 V_6 -16 W_2
    \right) \\
c_3 &=& {N-1\over (N-2)^3} \left\{
     - {1\over 384} (10 N^3 - 79 N^2 + 186 N - 137) +\,
       {1\over 192 \pi} (2 N -3) (3N^2 - 26 N + 37) \right. \nonumber \\
    && \quad \left. + {1\over 32\pi^2} (N-1)(3N-5) -
       {1\over 48 \pi^3} (3 N^3 - 18 N^2 + 38 N - 27)\right\} \nonumber \\
    && + {N-1\over N-2}\left\{ - {1\over 192 \pi^3} (2 N^2 + 57 N - 134)
      \zeta(3) - {1\over48} (N-6) \, J\right. \nonumber \\
    && \qquad + \left. {G_1\over 4\pi} \left[(5 N -9) \pi -
       2 (2 N - 3)\right]\right\} \nonumber \\
    && +\, (N-1)\left\{
       {1\over 2\pi} (3-\pi)\, R -
       {N-3\over 16} (K - 2 V_3) \right. \nonumber \\
    && \quad \left. + {N-7\over12} V_1 + {1\over2} (-2 V_4 + V_5 + V_6 +
       8 W_2) \right\}
\end{eqnarray}
A check of these expressions is provided by the $1/N$-expansion. In Appendix B
we have verified that these expressions are in agreement with the $1/N$ results
of \cite{Campostrini_90ab}.

Numerically  we have
\begin{eqnarray}
a_1 &\approx& {1\over N-2} (-0.0490 - 0.0141 N ) \\
a_2 &\approx& {1\over (N-2)^2} (0.0688 - 0.0028 N + 0.0107 N^2 - 0.0129 N^3) \\
b_1 &\approx& {1\over N-2} (-0.1888 + 0.0626 N ) \\
b_2 &\approx& {1\over (N-2)^2} (0.1805 - 0.0302 N - 0.0080 N^2 - 0.0108 N^3) \\
c_1 &\approx& {N-1\over N-2} \, 0.0908 \\
c_2 &\approx& {N-1\over (N-2)^2} (-0.0316 - 0.0120 N + 0.0149 N^2) \\
c_3 &\approx& {N-1\over (N-2)^3} (0.0198 + 0.0005 N - 0.0011 N^2 - 0.0102 N^3
              + 0.0041 N^4)
\end{eqnarray}

Using the results of \cite{CP-3loop} we can also compute the second analytic
correction to all non-derivative dimension-zero operators. A suitable basis
is given by
\be
{Y^{\alpha_1,\ldots,\alpha_n}_n} \, =\,
   \bsigma^{\alpha_1}\ldots \bsigma^{\alpha_n} - \,
   \hbox{`` Traces"}
\ee
where ``Traces" must be such that ${Y^{\alpha_1,\ldots,\alpha_n}_n}$
is completely symmetric and traceless. These polynomials are
irreducible $O(N)$-tensors of rank $n$ and thus they renormalize
multiplicatively with no off-diagonal mixing.
We define the spin-$n$ susceptibility as
\be
\chi^{(n)}=\, \sum_x \< {Y^{\alpha_1,\ldots,\alpha_n}_n}(0)\,
                        {Y^{\alpha_1,\ldots,\alpha_n}_n}(x) \>
\ee
Standard renormalization group arguments give \cite{BLZ_2}
\be
\chi^{(n)}=\,
    C_\chi^{(n)}\, e^{4\pi\beta/(N-2)} \,
   \left( {2\pi\beta\over N-2}\right)^{-[2 + n(N+n-2)]/(N-2)}\,
  \left[ 1 + \sum_{k=1}^\infty {b_k^{(n)}\over \beta^k} \right]
\ee
The non-universal constant $C_\chi^{(n)}$ cannot be estimated in
perturbation theory. A general expression is available only in
the large-$N$ limit. We have
\be
C_\chi^{(n)}\,=\, {(2\pi)^n\over32}\,
   \lim_{m_0\to0} m_0^2\int \prod_{i=1}^n {d^2p_i\over (2\pi)^2}\,
    (2\pi)^2 \delta\left( \sum_{i=1}^n p_i\right) \prod_{i=1}^n
    {1\over \hat{p}_i^2 + m_0^2}\, +\, O(1/N)
\ee
Explicitly $C^{(2)}_\chi = \pi/32$, $C^{(3)}_\chi = \pi^3 R/4$. For $n=2$
we also computed the first $1/N$ correction to get
\cite{CP-unosuN}
\begin{eqnarray}
C^{(2)}_\chi &=& {\pi\over32} \left[ 1 + {1\over N}\left(-2 - \pi
   + 6 \log{\pi\over4} - {36\over \pi^2} \zeta' (2) \right)\,
          +\, O(1/N^2) \right] \nonumber \\
       &\approx& 0.0982 \left(1 - 3.171/N + O(1/N^2) \right)
\end{eqnarray}
where $\zeta'(2) = -0.9375482$.
Let us now consider the analytic corrections. In \cite{CP-3loop} we considered
the ratio
\be
{\chi^{(n)}\over \xi^2}\, =\,
   R^{(n)} \left( {2\pi\beta\over N-2}\right)^{-n(N+n-2)/(N-2)}\,
  \left[ 1 + \sum_{k=1}^\infty {c_k^{(n)}\over \beta^k}\right]
\ee
and computed $c_1^{(n)}$ and $c_2^{(n)}$. Their expression is
\begin{eqnarray}
c_1^{(n)} &=& {1\over 4\pi} {n(N+n-2)\over N-2} (\pi-2) \\
c_2^{(n)} &=& {1\over2} \left(c_1^{(n)}\right)^2 +
             {n(N+n-2)\over N-2}
       \left[ (N-2)G_1 - {1\over 8\pi} - {1\over96} (3N-14)\right]
\end{eqnarray}
Then we get immediately
\begin{eqnarray}
b_1^{(n)} &=& 2 a_1 +\, c_1^{(n)} \\
b_2^{(n)} &=& 2 a_2 + a_1^2 + c_2^{(n)} + 2 a_1 c_1^{(n)}
\end{eqnarray}
Numerically
\begin{eqnarray}
b_1^{(n)} &=& {1\over N-2} \left( - 0.0980 - 0.0283 N + 0.0908\, n (n + N -2)
       \right)   \\
b_2^{(n)} &=& {1\over (N-2)^2}
      \left[ 0.1401 - 0.0042 N + 0.0215 N^2 - 0.0257 N^3
      \right. \nonumber \\
     && \qquad + \left. n(n+N-2)(-0.0363 - 0.0187 N + 0.0149 N^2)  \right.
      \nonumber \\ [3mm]
       && \qquad \left. + 0.0041 \, n^2 (n+N-2)^2 \right]
\end{eqnarray}
\begin{table}
\begin{center}
\tabcolsep 6pt      
\doublerulesep 2pt  
\begin{tabular}{|l||c|c||c|c|c|}
\hline
$N$ & $\beta$ & $\xi$ & ${\cal R}^{(2-loop)}$ & ${\cal R}^{(3-loop)}$ &
                        ${\cal R}^{(4-loop)}$  \\
\hline
3 & 1.85 & 89.16(21) & 0.743(2) & 0.781(2) & 0.830(2) \\
3 & 2.25 & 1049(7) & 0.861(6) & 0.897(6) & 0.934(6) \\
3 & 2.60 & 8569(92) & 0.901(10) & 0.934(10) & 0.962(10) \\
3 & 3.00 & 94.6(1.6) $\cdot 10^3$ & 0.930(16) & 0.959(16) & 0.981(17) \\
\hline
4 & 2.50 & 34.85(9) & 0.911(2) & 0.930(2) & 0.953(2) \\
4 & 2.80 & 86.07(37) & 0.927(4) & 0.945(4) & 0.964(4) \\
\hline
8 & 5.80 & 33.41(8)  & 0.985(2) & 0.990(2) & 0.995(2) \\
\hline
\end{tabular}
\end{center}
\caption{
Ratio of the Monte Carlo results for $\xi$ and the theoretical predictions
at two loops, three loops and four loops. $N=3$ data are taken from
\protect\cite{CEPS_O3_FSS}, $N=4,8$ from \protect\cite{Wolff_O4_O8}.
}
\label{MCcomparison}
\end{table}

We want now to compare our four-loop prediction with the available
Monte Carlo data for the correlation length. We define
\be
    {\cal R}^{(n-loop)}(\beta) \, =\, {\xi_{MC} (\beta)\over
         \xi_{th}^{n-loop} (\beta)}
\ee
where $\xi_{MC} (\beta)$ is the Monte Carlo value of the correlation length
\footnote{We consider here the isovector exponential correlation length
$\xi_V^{(exp)}$. Notice that the data in \cite{CEPS_O3_FSS} for $N=3$
refer instead to $\xi^{(2)}_V$. The two quantities differ however
by less than 0.1\% \cite{Meyer_ratio} and thus we will
ignore the difference.
} and
$\xi_{th}^{n-loop} (\beta)$ is the theoretical $n$-loop prediction
given by \reff{eq2.90} and \reff{Hasenfratzconst}. In table
\reff{MCcomparison} we report $\cal{R}$ for some selected values of
$N$ and $\beta$.
It is evident that the inclusion of the four-loop correction improves the
agreement between theory and ``experiment". The remaining discrepancy
at the largest $\beta$-values available today
is now 2\% for $O(3)$, 4\% for $O(4)$ and 0.5\% for $O(8)$
and it can presumably be ascribed to the neglected higher-loop corrections.

We want now to try to keep into account the higher-loop corrections using the
large-$N$ results of \cite{Campostrini_90ab}. Let us first define
\be
a_n^{(1/N)} (N) \, =\, N^{n-1} \bar{a}_n
\ee
The coefficients $a_n^{(1/N)}$ are the leading contribution to $a_n$ in the
limit $N\to\infty$ and $\bar{a}_n$ are numerical coefficients which can be
computed using the $1/N$ expansion. Their explicit value for
$n=1\ldots6$ is reported in Table \reff{tavolaunosuN}. For $n=1$ and $n=2$
we can compare $a_n^{(1/N)}$ with the exact value $a_n$. One finds
that in both cases the ratio $a_n/a_n^{(1/N)}$ is a decreasing function of
$N$ reaching the limiting value of one for $N\to\infty$. The convergence is
however slow. For $N=4$ we have indeed
\begin{eqnarray}
         {a_1(4)\over a_1^{(1/N)}(4)} &=& 3.73 \\
         {a_2(4)\over a_2^{(1/N)}(4)} &=& 2.88
\end{eqnarray}
while for $N=8$ we get
\begin{eqnarray}
         {a_1(8)\over a_1^{(1/N)}(8)} &=& 1.91 \\
         {a_2(8)\over a_2^{(1/N)}(8)} &=& 1.58
\end{eqnarray}
The approximation is good at the 10\% level only for $N\gtapprox 50$
(resp. $N\gtapprox 35$) for $a_1$ (resp. $a_2$).
Nonetheless we can try to use the $1/N$ results to get a rough idea
of the role of higher loop corrections.

$N=8$ should be the case where the approximation works better. In this case we
will assume that the
coefficients $a_n(N=8)$ for $n\ge 3$ are given by
\be
      a_n (N=8)\, =\, k_n\, 8^{n-1} \bar{a}_n
\ee
where $\bar{a}_n$ are defined in Table \reff{tavolaunosuN} and $k_n$ is a
number
that we will vary between about 1 and 2. In this way we can get an
estimate for ${\cal R}^{(8-loop)}$. For $\beta = 5.80$ (see Table
\reff{MCcomparison}) we obtain  for $k_n \equiv1$
${\cal R}^{(8-loop)}=0.998$; while for $k_n \equiv2$ we have
${\cal R}^{(8-loop)}=1.001$. The eight-loop correction is of order
$3 k_6 \cdot 10^{-4}$. Thus at this order we would expect an agreement
at the order of 0.1\% and this is indeed what we get from this rough
approximation.

We can try the same rough approximation for $N=4$ , writing, for $n\ge3$,
$a_n(N=4) = k_n 4^{n-1} \bar{a}_n$.
For $\beta = 2.80$ we get ${\cal R}^{(8-loop)}=0.970$ (resp. 0.994)
for $k_n \equiv1$ (resp. 5).
Although in this case it is very difficult to make
any quantitative statement this calculation shows that the numbers are in
the correct ball-park.

{}From this analysis we can thus conclude that the theoretical
prediction of \cite{Hasenfratz-Niedermayer_1,Hasenfratz-Niedermayer_2}
is in very good agreement with the Monte Carlo data.
We do not discuss here other long-distance quantities like the
vector and tensor susceptibility. A detailed comparison with
Monte Carlo data will appear elsewhere \cite{CEPS_93}.

\section*{Acknowledgments}
We thank Massimo Campostrini, Paolo Rossi and Alan Sokal
for many useful comments.

\appendix
\section{Technical Details}   \label{secA}

\subsection{Notations} \label{secA1}
In this Appendix we introduce the notations we have used in the
explicit computation of $w_3$ and $\gamma_3$. The one-loop
perturbative results will be written in terms of the integral
\be
I(h) =\, \int_{-\pi}^\pi {d^2p\over (2\pi)^2} \Delta(p)
\ee
where $\hat{p}_\mu = 2 \sin (p_\mu/2)$,
$\hat{p}^2 = \sum_\mu \hat{p}^2_\mu$ and
\be
\Delta(p) = {1\over \hat{p}^2 + h}
\ee
The integral $I(h)$ is logarithmically divergent for $h\to 0$. Explicitly
\be
I(h) = {2\over \pi (4 + h)} K\left( {4 \over 4 + h}\right) =\,
-{1\over 4\pi} \log \left( {h\over 32}\right) +\,
O(h \log h)
\ee
where $K$ is a complete elliptic integral of the first kind.
We will also use
\be
I_2 (h) =\, \int_{-\pi}^\pi {d^2p\over (2\pi)^2} \Delta(p)^2 =\,
   - {d I(h)\over d h}
\ee
We need also some basic two-loop and three-loop integrals.
To simplify the notation let us first introduce
\begin{eqnarray}
\hskip -50pt   d\nu (p,h) &=& {d^2q\over (2 \pi)^2}  \,
              {d^2r\over (2 \pi)^2}  \,
              {d^2s\over (2 \pi)^2}  \,
              (2 \pi)^2 \delta (p + q + r + s)
              \Delta(q) \Delta(r) \Delta(s) \\
\hskip -50pt   d\mu(h)    &=& {d^2q\over (2 \pi)^2}  \,
              {d^2r\over (2 \pi)^2}  \,
              {d^2s\over (2 \pi)^2}  \,
              {d^2 t\over (2 \pi)^2}  \,
              (2 \pi)^2 \delta (q + r + s + t)
              \Delta(q) \Delta(r) \Delta(s) \Delta(t)
\end{eqnarray}
We will use the following integrals which already appear in our previous
work:
\begin{eqnarray}
G_1 &=& - {1\over4} \int d\nu(0,0) \, \Delta(q)
       \left[\hat{q}^2 - \hat{r}^2 - \hat{s}^2\right]\,
       \sum_\mu \hat{q}^4_\mu \\
R   &=& \lim_{h\to 0} \, \left[h \int d\nu(0,h) \right]\\
J   &=& \int d\mu(0)
  \left(\sum_\mu \hat{q}_\mu \hat{r}_\mu \hat{s}_\mu \hat{t}_\mu\right)^2 \\
K   &=& \int d\mu(0) \,
            \left[(\widehat{q+r})^2 - \hat{q}^2 - \hat{r}^2\right]\,
            \left[(\widehat{s+t})^2 - \hat{s}^2 - \hat{t}^2\right] \\
L_1 &=& \int d\mu(0)\, {1\over \hat{q}^2 (\widehat{s+t})^2} \, \times
         \nonumber \\
    && \, \sum_{\mu\nu} \hat{q}_\mu \hat{q}_\nu \hat{s}_\mu \hat{s}_\nu
          \hat{t}_\mu \hat{t}_\nu
          \left[ \hat{r}_\mu \hat{r}_\nu (\widehat{s+t})^2 -
          (\widehat{s+t})_\mu (\widehat{s+t})_\nu \hat{r}^2\right]
\end{eqnarray}

We have moreover introduced a set of 8 new constants . The quantities
$V_1,\ldots V_6$ correspond to lattice infrared-finite integrals
and are explicitly given by
\begin{eqnarray}
V_1 &=& \int d\mu(0)
     \, \sum_\mu \hat{q}_\mu \hat{r}_\mu \hat{s}_\mu \hat{t}_\mu \\
V_2 &=& \int d\mu(0) \,
         {1\over \left[(\widehat{q+r})^2 \right]^2 \hat{q}^2} \,
   \left[ (\widehat{s+t})^2 - \widehat{s}^2 - \widehat{t}^2 \right]
   \, \times  \nonumber \\
    &&   \sum_\mu \left[ {1\over2}(\widehat{q+r})^2_\mu \hat{q}^2 +
        (\widehat{q+r})^2 \hat{q}_\mu^2 \right] \,
         \left[(\hat{q}_\mu^2 + \hat{r}^2_\mu)
                 (\widehat{q+r})^2 - (\hat{q}^2 + \hat{r}^2)
                 (\widehat{q+r})^2_\mu\right]    \nonumber \\ \\
V_3 &=& \int d\mu(0) \, {1\over \left[(\widehat{q+r})^2 \right]^2} \,
       \left[ (\widehat{s+t})^2 - \widehat{s}^2 - \widehat{t}^2 \right] \,
       \left[ (\widehat{q+r})^2 - \widehat{q}^2 - \widehat{r}^2 \right] \,
       \sum_\mu (\widehat{q+r})^4_\mu  \nonumber \\ \\
V_4 &=& \int d\mu(0) \, {1\over (\widehat{q+r})^2 } \,
       \sum_\mu \hat{r}^2_\mu \hat{t}^2_\mu \,
       \sum_\nu \sin q_\nu \sin s_\nu   \\
V_5 &=& \int d\mu(0) \, {1\over (\widehat{q+r})^2 } \,
        \sum_\mu \hat{q}_\mu \hat{r}_\mu \hat{s}_\mu \hat{t}_\mu \,
        \sum_\nu \sin^2 (q + r)_\nu  \\
V_6 &=& \int d\mu(0) \, {1\over (\widehat{q+r})^2 } \,
        \left[ (\widehat{s+t})^2 - \widehat{s}^2 - \widehat{t}^2 \right] \,
        \sum_\mu (\widehat{q+r})^2_\mu \hat{q}_\mu \hat{r}_\mu
                \cos \left( {q+r\over2}\right)_\mu
\end{eqnarray}
We introduce also $W_1$ and $W_2$ as the finite part of infrared-divergent
integrals:
\begin{eqnarray}
W_1 &=& \lim _{h\to0} \left[ \int d\nu(0,h) \Delta(q)
        \sum_\mu \left( \hat{q}^4_\mu - 2 \hat{q}^2_\mu \hat{r}^2_\mu \right)
      \, +\,
   {1\over2} I(h)^2 - {3\over 8\pi} I(h) \right]\\
W_2 &=& \lim_{h\to0} \left[\int d\mu(h) \, \Delta(q+r)
         \sum_{\mu\nu} \sin q_\mu \sin r_\nu \sin s_\mu \sin t_\nu
        \right.
        \nonumber \\
    && -\,  \left.
         {1\over6} I(h)^3 + \left( {1\over8} - {1\over 8\pi}\right) I(h)^2
          - \left( {1\over32} + {1\over 16 \pi^2} - {1\over 16 \pi}
          - {R\over2} \right) I(h) \right]
\end{eqnarray}
The numerical value of all the constants is reported in Table~\ref{tavola1}.

The numerical computation of $V_1,\ldots$, $V_6$ does not present any
difficulty as the integrals are infrared-finite. More tricky is
handling the integrals leading to $W_1$ and $W_2$. In this case
we have used a method which was
introduced in the context of the $1/N$ expansion in~\cite{CR_RNC} .
Let us consider
the case of $W_1$. First of all let us determine the divergent terms
which must be subtracted from the original integral.
If we introduce
\be
d_\mu (q,h)\, = \int {d^2r\over (2\pi)^2}\,
   \Delta(r) \Delta(q+r)\, (\widehat{q+r})_\mu \hat{r}_\mu
\ee
we can rewrite
\be
 \int d\nu(0,h) \Delta(q)
        \sum_\mu \left( \hat{q}^4_\mu - 2 \hat{q}^2_\mu \hat{r}^2_\mu \right)
  \, = \,
 -2 \int{d^2 q\over (2\pi)^2}\, \Delta(q)^2 \,
       \sum_\mu \hat{q}^2_\mu \cos {q_\mu\over2}\, d_\mu (q,h)
\label{intW1}
\ee
Then we determine the behaviour
of $d_\mu(q,h)$ for $q\to0$ by writing an
integral representation for $d_\mu (q,h)$. Using the technique
presented in~\cite{CR_RNC} we get
\begin{eqnarray}
d_\mu (q,h) &=& {1\over 2\pi} \cos{q_\mu\over2} \,
        \int^1_0 dx \, {1\over a_\mu^2}
       \left\{ {4 \zeta\over (a_1 a_2)^{1/2}} K(\zeta)\, + \right.
           \label{dmurepint}\\
      && \quad \left. \left[ a_\mu^2 (4 + h) - (4 + h)^2 +
          4 x(1-x) (2 \hat{q}^2_\mu - \hat{q}^2)\right]
          {\zeta^3\over (a_1 a_2)^{3/2}}\, {E(\zeta)\over 1 - \zeta^2}
         \right\} \nonumber
\end{eqnarray}
where $E(\zeta)$ and $K(\zeta)$ are complete elliptic integrals
\cite{Gradshteyn} and
\begin{eqnarray}
\zeta^2 &=& {4 a_1 a_2 \over (4 + h)^2 - (a_1 - a_2)^2} \\
a_\mu^2 &=& 4 \left[ 1 - x(1-x) \hat{q}^2_\mu\right]
\end{eqnarray}
Using \reff{dmurepint} we can now compute the expansion of
$d_\mu(q,h)$ for $q^2\to0$, $h\to 0$, with $q^2/h$ arbitrary. We get
\begin{eqnarray}
d_\mu^{(exp)} (q,h) &=& {1\over 8\pi} \,
       \left\{ 1 - \log{h\over 32} + {1\over \rho}
       \left(1 + {4 h\over q^2}\right)\, \log \left({\rho-1\over\rho+1}\right)
       \right. - \nonumber \\
      && \quad \left. {2 q^2_\mu \over q^2} \left[1 +
            {2 h \over q^2 \rho} \log \left( {\rho-1\over \rho+1}\right)
            \right]\right\}
\end{eqnarray}
where
\be
\rho =\, \sqrt{1 + {4 h\over q^2}} \label{A.26}
\ee
Then we rewrite \reff{intW1} as
\begin{eqnarray}
&& - 2 \int {d^2q\over (2\pi)^2}\, \left[
   {1\over (\hat{q}^2+h)^2}\sum_\mu \hat{q}^2_\mu \cos{q_\mu\over2}
    d_\mu (q,h)\, -\, {1\over (q^2 + h)^2} \sum_\mu q^2_\mu d_\mu^{(exp)} (q,h)
     \right] \, \nonumber \\
&& - 2 \int {d^2q\over (2\pi)^2}\,{1\over (q^2 + h)^2} \,
      \sum_\mu q^2_\mu d_\mu^{(exp)} (q,h)
\label{intdmuseparato}
\end{eqnarray}
The first integral is infrared finite. Thus we can take the limit $h\to 0$
obtaining
\be
- 2 \int {d^2q\over (2\pi)^2}\, \left[
     {1\over (\hat{q}^2)^2}\sum_\mu \hat{q}^2_\mu \cos{q_\mu\over2}
      d_\mu (q,0)\, -\, {1\over 8\pi} \left( {1\over q^2} -
      2 {\sum_\mu q^4_\mu \over (q^2)^3} - {1\over q^2} \log {q^2\over 32}
      \right) \right]
\ee
Although this integral is finite its numerical evaluation is complex as the
integrand is a difference of two divergent quantities. To get
stable results we have split the integration domain in two parts:
a disk $D_\epsilon$ of radius $\epsilon=0.1$ around the origin and the
remaining region $R_\epsilon = [-\pi,\pi]^2 - D_\epsilon$. The integration
over $R_\epsilon$ is done numerically; to compute the integral over
$D_\epsilon$ we have first expanded the integrand up to $O(q^{12})$
and then we have performed the integration analytically.
In the implementation a useful check is provided by the expression
of $d_\mu (q,0)$ along the diagonal, i.e. for $q=(l,l)$ which can be
computed exactly
\be
d_\mu ((l,l),0) =\, - {1\over 4\pi} \log \left| \tan {l\over4}\right|
\ee

Let us now compute the second integral in
\reff{intdmuseparato} which is still infrared divergent.
We first change the integration domain: if $\Lambda > \pi \sqrt{2}$
we rewrite it as
\begin{eqnarray}
&& - 2 \int_{D_\Lambda} {d^2q\over (2\pi)^2}\,{1\over (q^2 + h)^2}
      \sum_\mu q^2_\mu d_\mu^{(exp)} (q,h)\, \nonumber \\
&& + 2 \int_{D_\Lambda - [-\pi,\pi]^2} {d^2q\over (2\pi)^2}\,{1\over (q^2)^2}
      \, \sum_\mu q^2_\mu d_\mu^{(exp)} (q,0) \, + O(h)
\end{eqnarray}
The second integral is easily computed numerically, while the first gives
\be
   - {1\over 32 \pi^2} \log^2{h\over \Lambda^2} \,
   - {3\over 32 \pi^2} \log {h\over \Lambda^2}  \, -\,
     {1\over 32 \pi^2} - {R\over 2}
\ee
In the calculation the choice of $\Lambda$ is completely arbitrary.
We have chosen $\Lambda=\sqrt{32}$ as in this case it is simple to
replace $\log h$ with $I(h)$.

The calculation of $W_2$ is completely analogous. Introducing
\be
d_{\mu\nu} (q,h)\, =\, \int {d^2r\over (2\pi)^2} \,
   \Delta(r) \Delta(q+r) \sin r_\mu \sin (q + r)_\nu
\ee
we must compute
\be
\int {d^2q\over (2\pi)^2}\, \Delta(q)\, \sum_{\mu\nu} d_{\mu\nu}(q,h)
        d_{\mu\nu}(q,h)
\ee
First of all we compute
an integral representation for $d_{\mu\nu} (q,h)$. We get
\be
    d_{11} (q,h) =\, \cos{q_1\over2}\, d_1(q,h) -\,
                     {1\over4}\, \bar{d}_1 (q,h)
\ee
where $d_1(q,h)$ is given in \reff{dmurepint} and
\begin{eqnarray}
\bar{d}_1 (q,h) &=& {1\over2} - {1\over\pi} K\left( {4\over 4+h}\right) \,
     - \nonumber \\
  && {1\over\pi} \int^1_0 dx {E(\zeta)\over 1-\zeta^2}\,
    {\zeta^3 \over (a_1 a_2)^{3/2} } {1\over a_1^2} \,
    \left\{2 \cos^2 {q_1\over2}\left( (4 + h)^2 + a_1^2 - a_2^2\right) \right.
   \nonumber \\
  && \qquad \left. - \cos^2 {q_1\over2} (4 + h) (4 + a_1^2) -
            (1-2x)^2 (4+h) \left[\hat{q}^2_1 +
                    \hat{q}_2^2 {a_1\over a_2}
            {2-\zeta^2\over \zeta^2} \right]\right\} - \nonumber \\
  && {2\over\pi} \int^1_0 dx \, K(\zeta)
     {\zeta \over (a_1 a_2)^{1/2} } {1\over a_1^2} \,
     \left[ - 4 \cos^2 {q_1\over2} + {1\over a_2^2} (1-2x)^2 (4+h)
          \hat{q}^2_2\right]
\end{eqnarray}
and
\begin{eqnarray}
d_{12}(q,h) &=& - {4 + h\over \pi} \sin q_1 \sin q_2 \, \times \nonumber \\
     &&\int_0^1 dx\,  x(1-x) {\zeta \over (a_1 a_2)^{5/2} }
       \left[ {2-\zeta^2\over 1-\zeta^2} E(\zeta) - 2 K(\zeta)\right]
\end{eqnarray}
{}From these integral representations we thus get in the limit
$q^2\to 0$, $h\to 0$ with $q^2/h$ arbitrary
\begin{eqnarray}
d_{\mu\nu}^{(exp)} (q,h) &=&
  {1\over 8\pi} \delta_{\mu\nu}\left[ 2 - \pi - \log{h\over 32} +
  {1\over \rho} \left( 1 + {4 h\over q^2}\right)
  \log\left( {\rho-1\over \rho+1}\right)\right] \nonumber \\
  && \qquad - {1\over 4\pi} {q_\mu q_\nu \over q^2}\,
     \left[ 1 + {2 h\over q^2 \rho}
\log\left( {\rho-1\over \rho+1}\right)\right]
\end{eqnarray}
The computation is analogous to the previous case. The relevant integral
is now
\begin{eqnarray}
&& \int_{D_\Lambda} {d^2 q\over (2\pi)^2 } {1\over q^2 + h}
    \sum_{\mu\nu} d_{\mu\nu}^{(exp)} d_{\mu\nu}^{(exp)} \, = \nonumber \\
&& \quad - {1\over 384 \pi^3} \log^3 {h\over \Lambda^2}\,  -\,
           {1\over 128 \pi^2} \left( 1 - {1\over \pi}\right)
           \log^2 {h\over \Lambda^2}  \nonumber \\
&& \quad - {1\over 4\pi}\left({1\over 32} + {1\over 16 \pi^2} -
           {1\over 16 \pi} - {R\over 2}\right)
           \log {h\over \Lambda^2}\, +\,
           {R\over 8} \left(1 - {3\over \pi} \right) +
           {17 \over 576 \pi^3} \zeta(3) + \, O(h \log^3 h) \nonumber \\
\end{eqnarray}
where $D_\Lambda$ is a disk of radius $\Lambda$ around the origin.

In order to check the  manipulations of the extremely cumbersome expression for
$\bar{d}_1(q,h)$ we found very useful the expression of $d_1(q,h)$
for $q=(l,l)$ and $h=0$ given by
\be
\bar{d}_1 ((l,l),0) \, =\, {1\over2} - {1\over 2\pi} +
   {1\over 2\pi} {\sin^2 l/2\over \cos l/2} \log\left| \tan {l\over4}\right|
\ee

\subsection{Some Lattice Integrals}
In this section we report a list of integrals we have used in our computation.

Two-loop integrals:
\begin{eqnarray}
&& \int d\nu(p,h) \, \Delta(q) \Delta(p+q)^{-2} \, =  \\
   && \qquad I(h)^2 +\,
   p^2\left[ 2 \int d\nu(0) - {1\over 4} I(h)^2 -
     {3\over 16 \pi} I(h) + {1\over2} (R-W_1)\right]  + \, O(h^2,h p^2,p^4)
    \nonumber \\
&& \int d\nu(p,h) \, \Delta(q)  \Delta(p+r)^{-2} \, = \nonumber \\
   && \qquad I_2(h) - {1\over 4\pi} \left(I(h) - {1\over4}\right) +
      I(h)^2 - {1\over4} I(h) \, +\nonumber \\
   && \qquad {1\over h} p^2 \left[
             {1\over 4\pi}\left(I(h) - {1\over4} + {1\over 2\pi}\right) -
      {2 R\over3} + O(h) \right] + \, O(h,p^4) \\
&& \int d\nu(p,h)  \, \Delta(q)
       \left(2\, \Delta(p+q)^{-1}  \Delta(p+r)^{-1} +
                          \Delta(p+r)^{-1}  \Delta(p+s)^{-1}\right) \,
         = \nonumber \\
   && \qquad I_2(h) + 2 I(h)^2 + {1\over h} p^2
      \left[ {1\over2\pi} \left( I(h) - {1\over 4\pi}\right)
       + {2 R \over3}
        + O(h) \right] + \, O(h,p^4) \\
&& \int d\nu(p,h)  \, \Delta(p+q)^{-1} \, = \nonumber \\
    && \qquad I(h)^2 + p^2 \left[ {1\over3} \int d\nu(0)
          - {1\over 16 \pi} I(h) + {R\over 12} - {W_1\over6}\right]
    \, + \, O(h^2,h p^2,p^4) \\
&& \int d\nu(p,h)  \, \Delta(p+q)^{-2} \, = \nonumber \\
    && \qquad 2 I(h) - {1\over4} + h\left( - I(h)^2 + {1\over2} I(h)\right)
       \nonumber \\
    && \qquad
       + p^2 \left[ I(h)^2 + I(h) \left( {1\over \pi} - {1\over2}\right) -
        {1\over16} + 2 G_1 - 2 R\right]  + \, O(h^2 , h p^2, p^4)\\
&& \int d\nu(p,h)  \, \Delta(p+q)^{-1} \Delta(p+r)^{-1} \, = \nonumber \\
    && \qquad I(h) + p^2\left[ I(h)^2 - {1\over 2\pi} I(h) + {1\over48}
       - G_1 + R \right] + \, O(h^2,hp^2,p^4) \\
&& \int d\nu(0,h) \Delta(q) \sum_\mu \sin^2 q_\mu \, =\, \nonumber \\
    && \qquad {2\over3} \int d\nu(0) - {1\over8} I(h)^2 -
       {1\over 32\pi} I(h) +
       {1\over 12} (2 R - W_1) +\, O(h)
\end{eqnarray}

Three-loop integrals:
\begin{eqnarray}
&&\int d\mu(h)  \, \Delta(q+r)^{-1} = {4\over3} I(h)^3 - {V_1\over3}
           - {7\over 192\pi^3} \zeta(3)  + O(h) \label{A.45} \\
&&\int d\mu(h) \, \Delta(q+r)^{-2} =
          4 I(h)^2 - I(h) + K +\, \nonumber \\
&& \quad + h\left[ - {8\over3} I(h)^3 + 2 \left( 1 - {1\over \pi}\right) I(h)^2
   + \left( -8 G_1 + {1\over4} - {1\over \pi^2}\right) I(h)\right.
   \nonumber \\
&& \qquad \; \left. + {1\over 8 \pi} - {1\over 4 \pi^3} +
     {77\over 192 \pi^3} \zeta(3) - {2\over \pi} G_1 +
     {2 V_1\over3} + 2 V_2 \right] + \, O(h^2)
                                     \label{A.46} \\
&& \int d\mu(h)  \Delta(q+r)^{-1} \Delta(q+s)^{-1}  \, = \nonumber \\
   && \quad  2 I(h)^2 - {1\over24} - {K\over2} + {J\over6} + \nonumber \\
   && \qquad h\left[ - {4\over3} I(h)^3 + {1\over \pi} I(h)^2
   + \left( 4 G_1 - {1\over 12} + {1\over 2 \pi^2}\right) I(h)\right.
   \nonumber \\
   && \qquad \;
    \left. - {1\over 16 \pi} + {1\over 8 \pi^3} -
           {35\over 192 \pi^3} \zeta(3) + {G_1\over \pi} - V_2
    - {2\over3} L_1  \right]
\end{eqnarray}

\subsection{Analytic
         evaluation of one-loop integrals}

\newcommand{\B}{{\cal B}}
In this appendix we discuss the evaluation of the most general one-loop
integral. Let us introduce the notation
\be
\B(r;n_x,n_y) =
\int_{-\pi}^\pi {d^2 k\over (2 \pi)^2}  \, {
\hat{k}^{2 n_x}_x \hat{k}^{2 n_y}_y \over
(\hat{k}^2 + h)^r } \label{Bint}\ee
for $n_i\ge 0$, $r>0$, and $x,y$ all
different.
In the following when one of the arguments $n_i$ is zero it will be
omitted as argument of $\B$.

Using a technique we developed for four-dimensional integrals~\cite{CMP} all
these integrals can be reduced to a sum of $I(h)$ and $I_2(h)$.

\newcommand\BB{\tilde\B_\delta}

We will firstly generalize (\ref{Bint}) by considering the following
more general integrals
\be
\BB(r;n_x,n_y) =
\int_{-\pi}^\pi {d^2k\over (2 \pi)^2}  \, {
\hat{k}^{2 n_x}_x \hat{k}^{2 n_y}_y  \over
(\hat{k}^2+h)^{r+\delta} } \ee
where $r$ is a positive or negative integer and $\delta$ a real number which
is introduced in order to avoid singular cases at intermediate stages of the
computation and which will be set to zero at the end.

The first thing we want to show is that each integral $\BB(r;n_x,n_y)$
can be reduced through purely algebraic manipulations to a sum of integrals
of the same type with $n_x=n_y=0$.

Indeed the integrals $\BB$ satisfy the following recursion relations:
\begin{eqnarray}
\BB(r;1)        & = & {1\over 2} \,[\BB(r-1) - h \BB(r)]        \nonumber\\
\BB(r;m,1) & = &  \BB(r-1;m) - \BB(r;m+1) -h \BB(r;m)
\end{eqnarray}
which can be obtained by the insertion of the trivial identity
\be
\hat{k}^2 = \sum_{q=1}^2 \hat{k}^2_q
\ee
and by keeping into account the cases in which the index $q$ equals one of the
other indices.
Furthermore, when $m>1$ we can write
\be
{({\hat k}^2_w )^m \over (\hat{k}^2+h)^{n+\delta}} =
4 {({\hat k}^2_w )^{m-1} \over (\hat{k}^2+h)^{n+\delta}} +
2 {({\hat k}^2_w )^{m-2} \over n + \delta - 1} \sin k_w
{\partial \over \partial k_w} { 1 \over (\hat{k}^2+h)^{n+\delta-1}}
\ee
Then, integrating by parts, we obtain the recursion relation:
\begin{eqnarray}
\lefteqn{\BB(n;\ldots,m)} \label{rec2}
        \\ & = & {m-1\over (n+\delta-1)}
        \BB(n-1;\ldots,m-1) -
        {4m-6\over n+\delta-1} \BB(n-1;\ldots,m-2) \nonumber \\
        &       & +\, 4\, \BB(n;\ldots,m-1)
\end{eqnarray}
These relations allow to reduce every integral $\BB(r;n_x,n_y)$
to a sum of the form
\be
\BB(r;n_x,n_y) = \sum_{p=r-n_x-n_y}^{r-1} \BB(p)
a_p (\delta)
\ee
For $p\ge 1$, $\lim_{\delta\to 0}$ $a_p(\delta)$ is finite while
for $p<1$ $a_p(\delta)$ may behave as $1/\delta$ when $\delta$ goes to zero,
meaning that we need to compute $\BB(p)$ including terms of order $\delta$
when $p\leq0$.

Now let us show that all $\BB(p)$ can be expressed in terms of
$\BB(1)$ and $\BB(2)$.
Indeed let us consider the trivial identity
\be
\BB(r;1,1)\, -\, 2\BB(r+1;2,1) \, -h\,\BB(r+1;1,1)\, =\, 0
\ee
If we apply the reduction procedure to the integrals appearing in this
relation we get the following two relations
\begin{eqnarray}
\BB(q) &=& {4+h\over q^3} \left[ q (3 q^2 + 3q + 1) - (3 q+2) \delta\right]
           \BB(q+1) \, \nonumber \\
       && \quad + \, {q+1\over q^3} (32 + 24 h + 3 h^2) (-q (q+1) + 2 \delta)
          \, \BB(q+2) \label{recindietro} \\
       && \quad + {h(4+h) (8+h)\over q^3} \left[q (q+1) (q+2) -
           \delta (3 q + 4)\right]\, \BB(q+3)  + \, O(\delta^2)
     \nonumber \\
\BB(q) &=& {(q-3)^2\over (q-2) (q-1)} {1\over h (4 + h) (8+h)} \BB(q-3)
            \nonumber \\
       && \quad -\, {3 q^2 - 15 q + 19\over (q-1) (q-2)}
       {1\over h (8 + h)} \BB(q-2) \nonumber \\
       && \quad +\, {q-2\over q-1} {3 h^2 + 24 h + 32\over h (4 + h) (8 + h)}
         \BB(q-1) + O(\delta)
\end{eqnarray}
If we now apply the first relation for $q<0$ and the second for $q>2$ we
express every integral as
\be
\BB(r;n_x,n_y)\, = A(\delta) \BB(0)\, +\, B(\delta) \BB(1) \, +\,
       C(\delta) \BB(2) \, +\, O(\delta)
\ee
A careful analysis of the structure of the recursion \reff{recindietro}
shows that $B(\delta)$ and $C(\delta)$ are finite for $\delta\to0$.
As the l.h.s. is also finite for $\delta\to 0$ we get
that also $A(\delta)$ is finite in this limit. Thus we can set $\delta=0$
to get
\be
\B(r;n_x,n_y)\, =\, A(0) +\, B(0) I(h) +\, C(0) I_2(h)
\ee
As a final remark, notice that the whole procedure is completely algebraic
and can be easily implemented on a computer using a symbolic language.

\section{Comparison with the $1/N$ expansion}

In this appendix we want to compare our four-loop result with the
$1/N$ results of \cite{Campostrini_90ab}. In the perturbative limit
we have
\begin{eqnarray}
 \xi &=& \xi_0 +\, {1\over N} \xi_1 \,+\, O(1/N^2) \\
\chi &=& \chi_0 +\, {1\over N} \chi_1 \,+\, O(1/N^2)
\end{eqnarray}
with
\begin{eqnarray}
\xi_0 &=& {1\over \sqrt{32}} \exp \left( {2 \pi \beta\over N}\right) \\
\chi_0 &=& {N\over 32 \beta} \exp \left( {4 \pi \beta\over N}\right)
\end{eqnarray}
and \cite{Campostrini_90ab}
\begin{eqnarray}
\xi_1 &=& \xi_0 \left[ {4 \pi \beta\over N} - \log {4 \pi \beta \over N}
       - 2 \log 2 - \gamma_E + 1 - {\pi\over2} \right] \nonumber \\
   && \qquad - 2 \pi \xi_0 \left[
       \int_{lat} {d^2 k\over (2\pi)^2 } {1\over \beta/N + A_0 (k)}
       \left( A_1(k) - B_1(k) A_0 (k)\right) \right. \nonumber \\
   && \qquad \qquad \left. - 2 \int_{k^2<32} {d^2 k\over (2\pi)^2 }
      {1\over k^2} \, {1\over \log(k^2/32) + 4 \pi \beta/N} \right]
     \label{xiunosuN}\\[3mm]
\chi_1 &=& \chi_0 \left[
        {8 \pi \beta\over N} - 3 \log {4 \pi \beta \over N}
       - 4 \log 2 - 3 \gamma_E +3 \gamma_C + 2 - \pi \right] \nonumber \\
    && \qquad - 4 \pi \chi_0 \left[
       \int_{lat} {d^2 k\over (2\pi)^2 } {1\over \beta/N + A_0 (k)}
       \left( A_1(k) - B_1(k) A_0 (k) -  {1\over 4\pi} B_1(k) -
       {1\over 4\pi \hat{k}^2} \right) \right. \nonumber \\
   && \qquad \qquad \left. - 3 \int_{k^2<32} {d^2 k\over (2\pi)^2 }
      {1\over k^2} \, {1\over \log(k^2/32) + 4 \pi \beta/N} \right]
     \label{chiunosuN}
\end{eqnarray}
where
\begin{eqnarray}
A_0 (k) &=& {1\over2} \int {d^2 q\over (2\pi)^2 }
            { \hat{k}^2 - \hat{q}^2 - (\widehat{k+q})^2 \over
             \hat{q}^2 (\widehat{k+q})^2} \\
A_1(k) &=& {1\over 2} \int {d^2 r \over (2\pi)^2 } {d^2 s \over (2\pi)^2 }\,
           (2 \pi)^2 \delta (k-r-s)
          {1\over \left( \hat{r}^2 \hat{s}^2\right)^2}
           \nonumber \\
       && \qquad \qquad \left[ - \hat{k}^2 (\hat{r}^2 + \hat{s}^2) +
               (\hat{r}^2)^2 + (\hat{s}^2)^2 - B_1(k) \hat{r}^2 \hat{s}^2
              (\hat{r}^2 + \hat{s}^2)\right]
\end{eqnarray}
and
\be
B_1(k) = \, - {2\over (\hat{k}^2)^2} \sum_\mu \sin^2 k_\mu -\, {1\over4}
\ee
A point must be noted in the solution \reff{xiunosuN}, \reff{chiunosuN}:
both $\xi_1$ and $\chi_1$ are expressed in terms of differences
of two integrals which have a non-integrable singularity for $k=0$.
This notation is however symbolic and it must be interpreted
in the following way: for two functions $f(k)$ and $g(k)$
we define
\be
\int_{lat} {d^2k\over (2\pi)^2} f(k) - \int_{k^2<32} {d^2k\over (2\pi)^2} g(k)
\, =\,  \int_B {d^2k\over (2\pi)^2} (f(k) - g(k)) -
        \int_C {d^2k\over (2\pi)^2} g(k)
\ee
where $B$ is the domain $[-\pi,\pi]^2$, $C = D_{\sqrt{32}} - B$ and
$D_r$ the disk centered in the origin of radius $r$. It is easily checked
that with this definition everything is well-defined for $k=0$ as
in both cases the singularity cancels in the difference.
{}From \reff{xiunosuN} and \reff{chiunosuN} we can easily derive
the large-$N$ behaviour of $a_2$, $b_2$ and $c_3$. We get
\begin{eqnarray}
{1\over N} a_2 &=& - 2 \pi \left[ \int_{lat} {d^2k\over (2\pi)^2}
               A_0(k) \left( A_0(k) B_1(k) - A_1(k)\right)
              \right. \nonumber \\
    && \qquad\left. + {1\over 8\pi^2} \int_{k^2<32} {d^2k\over (2\pi)^2}
   {1\over k^2} \log {k^2\over 32}\right] +\, O(1/N)
   \label{B.11} \\
{1\over N} b_2 &=& - 4 \pi \left[ \int_{lat} {d^2k\over (2\pi)^2}
               A_0(k) \left( A_0(k) B_1(k) - A_1(k)
                  + {1\over 4\pi} B_1(k) + {1\over 4\pi \hat{k}^2}\right)
              \right. \nonumber \\
    && \qquad\left. + {3\over 16\pi^2} \int_{k^2<32} {d^2k\over (2\pi)^2}
   {1\over k^2} \log {k^2\over 32}\right] +\, O(1/N)
   \label{B.12} \\
{1\over N^2} c_3 &=&  \int_{lat} {d^2k\over (2\pi)^2}
               A_0(k)^2 \left( B_1(k) + {1\over \hat{k}^2}\right)
               \nonumber \\
    && \qquad  + {1\over 16\pi^2} \int_{k^2<32} {d^2k\over (2\pi)^2}
   {1\over k^2} \log^2 {k^2\over 32} +\, O(1/N)  \label{B.13}
\end{eqnarray}
In order to compare these expressions with our results we must reexpress
these integrals in terms of our basic constants.
In the Appendix F of \cite{CR_RNC}~\footnote{Notice the change in notation:
$A_\#^{(\alpha)}=2 A_\#$, $B_\#^{(\alpha)}=2 B_\#$}
 one can find the basic relations
which are needed to compare the results of the $1/N$ expansion
with the standard three-loop result. As we will use them in the following
we report them here:
\begin{eqnarray}
&& \int_{lat} {d^2k\over (2\pi)^2} {1\over \hat{k}^2} \, -\,
   \int_{k^2<32} {d^2k\over (2\pi)^2} {1\over k^2} \, =\, 0 \label{unosuk}\\
&& \int_{lat} {d^2k\over (2\pi)^2} {A_0(k)\over \hat{k}^2} \, -\,
   \int_{k^2<32} {d^2k\over (2\pi)^2} {1\over 4\pi k^2} \log {k^2\over 32}
   \, =\, 0 \label{unosukA0} \\
&& \int_{lat} {d^2k\over (2\pi)^2} A_1(k) \, +\,
   \int_{k^2<32} {d^2k\over (2\pi)^2} {1\over 2\pi k^2}
         \left(\log {k^2\over 32} - 1\right)
   \, =\ - {1\over 8\pi^2} \\
&& \int_{lat} {d^2k\over (2\pi)^2} B_1(k) A_0(k) \, +\,
   \int_{k^2<32} {d^2k\over (2\pi)^2} {1\over 2\pi k^2} \log {k^2\over 32}
   \, =\ {1\over 32} - G_1
\end{eqnarray}
We must now derive analogous expressions for the integrals appearing in
$a_2$, $b_2$ and $c_3$.

Let us begin by considering the quantity
\be
\int_{lat} {d^2k\over (2\pi)^2} {A_0(k)^2\over \hat{k}^2} \, -\,
{1\over 16 \pi^2} \int_{k^2<32} {d^2k\over (2\pi)^2} {1\over k^2}
           \log^2 {k^2\over 32}
\ee
In this case we start from the three-loop integral
\be
T\, =\, \int d\mu(h) \Delta(q+r)^{-1}
\ee
whose value is reported in \reff{A.45}.
It can also be computed using the
method we presented in Appendix A.1 when we considered $W_1$ and $W_2$.
If we define
\be
d_0(q,h) \, =\, {1\over2} \int {d^2r\over (2\pi)^2} \Delta(r) \Delta(r+q)
\ee
we can rewrite $T$ as
\be
T =\, 4 \int {d^2q\over (2\pi)^2} \Delta(q)^{-1} d_0(q,h)^2
\ee
Now for $q\to 0$, $h\to 0$, $q^2/h$ fixed, $d_0(q,h)$ becomes its continuum
counterpart
\be
d_0^{exp} (q,h) \, =\, {1\over 4\pi q^2 \rho}
       \log \left( {\rho+1\over \rho-1} \right)
\ee
where $\rho$ is defined in \reff{A.26}. Thus we rewrite
\begin{eqnarray}
T &=& 4 \left\{ \int_{lat} {d^2k\over (2\pi)^2} \Delta(k)^{-1} d_0(k,h)^2
              - \int_{k^2<32} {d^2k\over (2\pi)^2} (k^2 + h)
                 d_0^{exp} (k,h)^2 \right\} \nonumber \\
  && + 4 \int_{k^2<32} {d^2k\over (2\pi)^2} (k^2 + h)
                 d_0^{exp} (k,h)^2
\end{eqnarray}
The second integral is easily done and gives
\be
\int_{k^2<32} {d^2k\over (2\pi)^2} (k^2 + h)
                 d_0^{exp} (k,h)^2 =\,
  - {1\over 192 \pi^3} \log^3 {h\over 32} - \, {5\over 256\pi^3} \zeta(3)
\ee
For the term in curly brackets let us notice that
\be
\Delta(k)^{-1} d_0(k,h)^2 - (k^2 + h) d_0^{exp} (k,h)^2
\ee
is integrable for $k\to0$ for all values of $h$. Thus we can expand the
integrand for $h\to 0$. Using the fact that for $h\to 0$ at $k$ fixed we have
\be
d_0 (k,h) \, =\, {1\over \hat{k}^2} \left( I(h) + A_0 (k)\right) +\,
       O(h \log h)
\ee
and the relations \reff{unosuk}/\reff{unosukA0}, we get at the end
\begin{eqnarray}
T & = & {4\over3} I(h)^3 -\, {5\over 64 \pi^3} \zeta(3) \nonumber \\
  && \qquad + 4 \left\{ \int_{lat} {d^2k\over (2\pi)^2}
        {A_0(k)^2\over \hat{k}^2} -\,  {1\over 16\pi^2}
      \int_{k^2<32} {d^2k\over (2\pi)^2} {1\over k^2} \log^2 {k^2\over32}
      \right\}
\end{eqnarray}
Comparing with \reff{A.45} we get
\be
\int_{lat} {d^2k\over (2\pi)^2}
        {A_0(k)^2\over \hat{k}^2} -\,  {1\over 16\pi^2}
      \int_{k^2<32} {d^2k\over (2\pi)^2} {1\over k^2} \log^2 {k^2\over32}
 \, =\, {1\over 96\pi^3} \zeta(3) -\, {V_1\over12}
\label{unosukA02}
\ee
Simple algebraic manipulations give also
\be
\int_{lat} {d^2k\over (2\pi)^2} B_1(k) A_0(k)^2 + \,
   {1\over 8\pi^2} \int_{k^2<32} {d^2k\over (2\pi)^2}
   {1\over k^2} \log^2 {k^2\over32} \, =\,
  - {1\over 48\pi^3} \zeta(3) - {K\over16}+ {V_1\over6} + {V_3\over8}
\label{B1A02}
\ee
The last integral which is needed is
\be
\int_{lat} {d^2k\over (2\pi)^2} A_1(k) A_0(k) + \,
   {1\over 8\pi^2} \int_{k^2<32} {d^2k\over (2\pi)^2}
   {1\over k^2} \left( \log {k^2\over32} - 1\right) \log{k^2\over32}
\ee
This quantity can be handled exactly in the same way starting
from
\be
\int d\mu(h) \Delta(q+r)^{-2} \Delta(q) =\,
-2 \int {d^2q\over (2\pi)^2} \Delta(q)^{-2} d_0(q,h)
           {\partial d_0 (q,h)\over \partial h}
\ee
which can be computed by derivation with respect to $h$ from \reff{A.46}.
The relevant integral is
\begin{eqnarray}
&& \hskip -60pt \int_{q^2<32} {d^2q\over (2\pi)^2} \Delta(q)^{-2}
d_0^{exp}(q,h)
           {\partial d_0^{exp} (q,h) \over \partial h}  = \nonumber \\
&&\hskip -60 pt \quad {1\over 2 \pi^3 h} \left( \log {h\over 32} + 1\right) +
  {1\over 32\pi^3}\left( {1\over3} \log^3 {h\over 32} +  \log^2{h\over 32}
  +  \log {h\over 32} - 1\right) + {41\over 512\pi^3} \zeta(3)
\end{eqnarray}
We finally get
\begin{eqnarray}
&& \int_{lat} {d^2k\over (2\pi)^2} A_1(k) A_0(k) + \,
   {1\over 8\pi^2} \int_{k^2<32} {d^2k\over (2\pi)^2}
   {1\over k^2} \left( \log {k^2\over32} - 1\right) \log{k^2\over32}
   \nonumber \\
&& \qquad =\, - {1\over 48 \pi^3} \zeta(3) + {1\over 64\pi} - {G_1\over 4\pi}
   + {V_1\over 6} + {V_2\over 4}
\label{A1A0}
\end{eqnarray}
We have checked numerically \reff{unosukA02}, \reff{B1A02} and \reff{A1A0}.
This provides also a check of the integrals \reff{A.45} and \reff{A.46}.

Using \reff{B.11}, \reff{B.12} and \reff{B.13}
we finally get for the coefficients $a_2$, $b_2$ and $c_3$
\begin{eqnarray}
{a_2\over N} &=& {1\over 32} - {G_1\over2} + {\pi\over8}
     \left( K + 4 V_2 - 2 V_3\right)  + \, O(1/N)\\
{b_2\over N} &=& {1\over32} + {\pi\over4}
     \left(K + 4 V_2 - 2 V_3\right) +\, O(1/N) \\
{c_3\over N^2} &=&   - {1\over 96\pi^3} \zeta(3) - {K\over16} +
      {V_1\over 12} + {V_3\over 8} +\, O(1/N)
\end{eqnarray}
{}From these expressions, using the
renormalization group relations \reff{eq2.90}/\reff{eq2.131}, we can compute
the large-$N$ contribution to $w_3^{latt}$ and $\gamma_3^{latt}$.
These expressions agree with \reff{eq2.31}/\reff{eq2.32}.

{}From the general expressions \reff{xiunosuN} and \reff{chiunosuN}
it is also possible to compute
the value of higher-loops coefficients. Defining
\begin{eqnarray}
\bar{a}_n &=& {a_n\over N^{n-1}} \label{abar}\\
\bar{b}_n &=& {b_n\over N^{n-1}} \label{bbar}
\end{eqnarray}
we report their values in Table\reff{tavolaunosuN}. We can also compute,
in view of the possibility of using {\em improved } expansions, the
perturbative expansion of the isovector energy
$E_V = \< \bsigma_0\cdot\bsigma_1\>$.
{}From \cite{Campostrini_90ab} we get
\begin{eqnarray}
E_V &=& 1 - {N-1\over 4\beta} - {1\over 4 N}\int {d^2 k\over (2\pi)^2}
         {1\over \beta/N + A_0 (k)} \nonumber \\
    &=& 1 - {N-1\over 4\beta}- {N\over 32\beta^2}
      + {1\over N} \sum_{n=3}^\infty
        \bar{e}_{n-2} {N^n\over \beta^n}
\end{eqnarray}
\begin{table}
\begin{center}
\tabcolsep 6pt      
\doublerulesep 2pt  
\begin{tabular}{|l||c|c|c|c|}
\hline
$n$ & $\bar{a}_n$  & $\bar{b}_n$ & $\bar{e}_n$ & $\bar{f}_n$ \\
\hline
1 & $-$ 0.01413 & \hphantom{$-$} 0.06259 & $-$ 0.005993 & 0.13650 \\
2 & $-$ 0.01289 & $-$ 0.01079 & $-$ 0.001568 & 0.02651 \\
3 & $-$ 0.00404 & $-$ 0.00401 & $-$ 0.000530 & 0.00929 \\
4 & $-$ 0.00145 & $-$ 0.00146 & $-$ 0.000221 & 0.00411 \\
5 & $-$ 0.00063 & $-$ 0.00063 & $-$ 0.000109 & 0.00212 \\
6 & $-$ 0.00031 & $-$ 0.00030 & $-$ 0.000063 & 0.00127 \\
\hline
\end{tabular}
\end{center}
\caption{
Values of the perturbative coefficients in the large-$N$ limit.}
\label{tavolaunosuN}
\end{table}
We can thus rewrite for the correlation length
\be
\xi\, = \bar{C}_{\xi}
\left({N-2\over 2\pi\beta_E}\right)^{1/(N-2)}
\exp\left( {2\pi\beta_E\over N-2}\right) \left(1 + {1\over N}
\sum_{n=1}^\infty \bar{f}_n{N^n\over\beta_E^n} \right)
\ee
where $\beta_E = (N-1)/(4 (1 - E_V))$ and
\be
\bar{C_\xi} =\,  C_\xi \exp\left({\pi\over 4(N-2)}\right)
\ee
The coefficients $\bar{e}_n$ and $\bar{f}_n$ are reported in Table
\reff{tavolaunosuN}.

\end{document}